\documentclass[10pt, conference, compsocconf]{IEEEtran}
\IEEEoverridecommandlockouts

\makeatletter
\def\ps@IEEEtitlepagestyle{
  \def\@oddfoot{\mycopyrightnotice}
  \def\@evenfoot{}
}
\def\mycopyrightnotice{
  {\footnotesize
  \begin{minipage}{\textwidth}
  \centering
  Copyright~\copyright~2018 IEEE. Personal use of this material is permitted. However, permission to use this  \\ 
  material for any other purposes must be obtained from the IEEE by sending a request to pubs-permissions@ieee.org.
  \end{minipage}
  }
}

\usepackage{cite}
\usepackage{amsmath,amssymb,amsfonts}
\usepackage[utf8]{inputenc}
\usepackage{algorithmic}
\usepackage{graphicx}
\usepackage{hyperref}
\usepackage{listings, multicol}
\usepackage{subcaption}

\usepackage{ulem}

\usepackage{todonotes}
\usepackage{textcomp}
\def\BibTeX{{\rm B\kern-.05em{\sc i\kern-.025em b}\kern-.08em
    T\kern-.1667em\lower.7ex\hbox{E}\kern-.125emX}}
\begin{document}

\title{A high-level C++ approach to manage local errors, %\sout{asynchronicity}
asynchrony and faults in  an MPI application\\
\thanks{This work has been supported in part by the German Research Foundation (DFG) through the Priority Programme \lq Software for Exascale Computing (SPP 1648)\rq, grant GO-1758/2 and EN-1042/2-2, the German Excellence Initiative through EXC 310 SimTech, and the German Federal Ministry of Education and Research (BMBF) through the research project HPC\textsuperscript{2}SE, grant 01/H16003A.}
}

\author{\IEEEauthorblockN{1\textsuperscript{st} Christian Engwer}
\IEEEauthorblockA{\textit{Applied Mathematics} \\
  \textit{University of Münster}\\
  Orleansring 10\\
  48149 Münster, Germany}
\and
\IEEEauthorblockN{2\textsuperscript{nd} Mirco Altenbernd}
\IEEEauthorblockA{\textit{IANS} \\
\textit{University of Stuttgart}\\
Allmandring 5b\\
70569 Stuttgart, Germany}
\and
\IEEEauthorblockN{3\textsuperscript{rd} Nils-Arne Dreier}
\IEEEauthorblockA{\textit{Applied Mathematics} \\
  \textit{University of Münster}\\
  Orleansring 10\\
  48149 Münster, Germany}
\and
\IEEEauthorblockN{4\textsuperscript{th} Dominik Göddeke}
\IEEEauthorblockA{\textit{IANS} \\
\textit{University of Stuttgart}\\
Allmandring 5b\\
70569 Stuttgart, Germany}
}

\maketitle

\begin{abstract}
  C++ advocates exceptions as the preferred way to handle
  unexpected behaviour of an implementation in the code. This does not integrate
  well with the error handling of MPI, which more or less always results in program
  termination in case of MPI failures. In particular, a local C++ exception can currently lead to a
  deadlock due to unfinished communication requests on remote hosts.
  At the same time, future MPI implementations are expected to include an API to
  continue computations even after a hard fault (node loss), i.e. the worst possible unexpected behaviour.

  In this paper we present an approach that adds extended exception propagation
  support to C++ MPI programs. Our technique allows to propagate local
  exceptions to remote hosts to avoid deadlocks, and to map MPI failures on
  remote hosts to local exceptions. A use case of particular interest are
  asynchronous \lq local failure local recovery\rq\ resilience approaches.
  Our prototype implementation uses MPI-3.0 features only. In addition we present a dedicated implementation, which integrates seamlessly
  with MPI-ULFM, i.e. the most prominent proposal for extending MPI towards fault tolerance.

  Our implementation is available at \url{https://gitlab.dune-project.org/christi/test-mpi-exceptions}.
\end{abstract}

\section{Introduction}
\label{sec:intro}

C++ programs comprise a wide range of target machines with many different architectures. Despite careful debugging, it is always possible that a program behaves unexpectedly in some way.
This is particularly important in the case of software frameworks, which we are mostly interested in: Packages like DUNE~\cite{Dune24:2016}, deal.II~\cite{dealII85:2017} and Trilinos~\cite{Heroux:2005:AOO} have a broad user base for PDE-related computations, and all use C++ template metaprogramming to relieve the user of the burden to implement common features over and over again.  Furthermore parallel software frameworks often hide at least coarse-grained parallelism from their users. In parallel numerical algorithms, unexpected behaviour can occur quite frequently: A solver could diverge, the input of a component (e.g. the mesher) could be inappropriate for another component (e.g. the discretiser), etc.

A well-written code should detect unexpected behaviour and provide the user with a possibility to react appropriately in their own code, instead of simply terminating with some error code.
For C++, \emph{exceptions} are the recommended method to handle this. With well placed exceptions and corresponding \emph{try-catch} blocks, it is possible to accomplish a more robust program behaviour. This holds both for framework developers and framework users.

For large-scale computations, MPI (\lq message passing interface\rq) is the de-facto standard for coarse-grained communication. The current MPI specification~\cite{mpi:2015} does not define any way to propagate exceptions from one so-called rank (process) to another. In the case of unexpected behaviour within the MPI layer itself, MPI programs simply terminate, maybe after a time-out.
This is a design decision that unfortunately implies a severe disadvantage in C++, when combined with the ideally asynchronous progress of computation and communication: An exception that is thrown locally by some rank can currently lead to a communication deadlock, or ultimately even to undesired program termination. Even though exceptions are technically an illegal use of the MPI standard (a peer no longer participates in a communication), it undesirably conflicts with the C++ concept of error handling.

With increasing degrees of parallelism and architectural complexity, it becomes even harder for framework developers to predict eventual misbehaviour and to provide appropriate infrastructure. When not resorting to C++ exceptions or synchronous global checkpoint-restart techniques, it is possible (albeit cumbersome and undesirable) with basic MPI features that a local process communicates its information to other participants to restore the functionality of an algorithm.
This does not hold however for failures within the MPI layer itself: The reliability of hardware is expected to become a non-negligible problem on upcoming extreme-scale systems~\cite{Dongarra:2014:AMR,Snir:2014:AFI}, and predictions of the Mean-Time-Between-Failure (i.e. the expected time span between two failures) hint at this problem to become the norm rather than the exception.
This makes it necessary to include support for failure propagation and thus fault-mitigation and fault-tolerance, for the full range from locally recoverable \lq irregularities\rq, the unrecoverable failure of complete ranks and their associated data, and anything in between.

\paragraph{Contribution}
We are convinced that any kind of unexpected behaviour in MPI-C++ programs should be treated and is treatable in the same way, i.e. through C++ exceptions. In this paper, we present a possible approach along with a prototype implementation to realise this claim. We follow C++11 techniques, e.g. use future-like abstractions to handle asynchronous communication.

\paragraph{State of the art} % (fold)
\label{par:state_of_the_art}

Current implementations of the MPI standard like MPICH, OpenMPI or IntelMPI do not provide an easy-to-use way to propagate unexpected behaviour from one process/rank to another: The standard does not mandate this, not even for failures like a node loss within the MPI layer itself. Current MPI implementations thus typically terminate (or deadlock) in such a situation. The most prominent proposal which suggests a suitable extension to the MPI standard currently is \emph{User-Level Failure Mitigation (ULFM)}~\cite{Bosilca:2016:FDP:3014904.3014941,Bland:2012:ULFM}. It allows users to define a workaround for the node loss scenario, e.g. clear the broken communicator and create a new one with a reduced number of processors, or include some spare nodes. This extension will provide a good solution for most of the arising problems, but it is still far away from being available on current HPC systems: ULFM might be included in the standard only from MPI-4 onwards. The approach we suggest works without ULFM, but includes a dedicated code path for any future MPI version that includes ULFM. We describe our ULFM integration in section~\ref{sub:ulfm_adoption}.

Furthermore the C++ API for MPI was dropped from MPI-3 since it offered no real advantage over the C bindings, instead of being a simple wrapper layer. MPI users coding in C++ are still using the C bindings, writing there own C++ interface/layer or using existing interfaces like Boost.MPI~\cite{boost:mpi:2017}. Although the Boost.MPI documentation states that the library \lq[\ldots] provides an alternative C++ interface to MPI that better supports modern C++ development styles, including complete support for user-defined data types and C++ Standard Library types and arbitrary function objects for collective algorithms\ldots\rq, there is currently no support for neither exception propagation nor fault-tolerance.

% paragraph state_of_the_art (end)

\paragraph{Use cases for our technique} % (fold)
\label{par:goal}

Due to the current discrepancy of supported software on different machines, which most of the time do not have MPI implementations with ULFM support, one goal of this work is to provide a flexible C++ interface for unified exception propagation. The infrastructure we provide allows to manage all kinds of local misbehaviour and most faults at the MPI level, in a C++ conforming way. Our approach is future-proof in the sense that it should easily enable switching to an ULFM-enhanced MPI version without the necessity to substantially change the user code, once such an MPI library becomes available. Switching to such an MPI deployment furthermore extends the type of faults which can be handled.

Our currently implemented prototype interface handles all faults which we describe in section \ref{sub:faults_and_failures} if the MPI installation supports ULFM. Otherwise only soft faults and thus exception propagation are supported.

We particularly focus on the quite general problem of propagating \emph{a local error to other MPI ranks}. This general problem is relevant in different scenarios and fault-tolerance concepts:
\begin{enumerate}%[label=\alph*),nosep]

  \item \emph{Local failure local recovery} (LFLR)~\cite{Teranishi:2014:TLF} techniques are based on recomputation of lost information, rather than resorting to a global checkpoint. For instance, Huber et al.~\cite{huber:2016:multigrid} or Göddeke et al.~\cite{Goeddeke:2015:FTF} pursue this idea for multigrid. If a node crashes, the lost approximation is recomputed, which mandates to re-establish the (lost) communicator.
  % \item Certain faults cannot be fixed purely locally, but require a \emph{roll-back in a small neighbourhood}. \todo{@Mirco: cite dafür bitte}In this case one would need some hierarchic escalation strategy to propagate the error to the neighbourhood in an efficient way, without impacting unaffected processes too much.

  %\emph{Note: }We will not consider this scenario within this work. This is mainly due to the fact that it is strongly connected to the topology of the machine and the algorithms. Therefore it has to be analysed for each algorithm and the communication has to be optimised by MPI itself. Nevertheless our interface should be easily extendable to such scenarios.
  \item In some scenarios it can be possible that failures are repairable in a sufficient way for the local computation but nevertheless would lead to bad behaviour on a more global level. This necessitates a \emph{local repair and (semi-) global action}. In this case one would need some hierarchical escalation strategy to propagate the error to the neighbourhood in an efficient way, without impacting unaffected processes too much.
  A prominent example are Krylov-type solvers, where a small local inconsistency can lead to a globally skewed Krylov space and thus to deteriorated convergence rates or even convergence to a wrong global solution.\label{example_krylov} This is in some sense comparable to the LFLR concept: A local repair and a global reset of the solver is sufficient to maintain a good convergence behaviour without a global rollback. This concept can drastically reduce the amount of necessary communication and possibly provides more efficient recovery for highly parallel systems.
  \item In the worst case we need a \emph{global roll-back} within the whole communicator. This means that we have to send a signal from possibly one rank to all ranks in the communicator, stop the current operation and go back to the state of the last checkpoint.
\end{enumerate}

% \todo{DG@Mirco: raffe den Sinn des folgenden Abschnitts nicht} The described concepts are all more or less related to the concept of \emph{algorithm-based fault-tolerance} (ABFT). This concept was already suggested by Huang and Abraham \cite{Huang:1984:ABF} in 1984 for dense matrix products. In general this ABFT mechanisms are embedded directly into the algorithm. If these mechanism detect an unusual behaviour they apply a repair method to reverse the fault or at least to reduce its effect.
%For highly parallel computation the, in contrast to the original concept, it could be necessary to propagate the information to other processes and combine the ABFT approach with a partial roll-back, restart or similar.
% paragraph goal (end)

%\paragraph{Structure} \todo{DG@all: würde ich weglassen}
%\label{par:structure}
%At first we will give some information on the necessity of fault-tolerance and the type of expected faults and the ones we will treat with our concept. Therefore we especially differentiate between two types of faults. Afterwards we describe the current relation between the ULFM concept and MPI. This will be followed up by a short description of our interface implementation. In particular we describe that this interface is a convenient solution to provide the user a way to manage local errors and faults without considering the supported MPI version for different machines. It follows a short performance analysis on a small cluster which compares the performance between different MPI libraries with and without ULFM support.
% paragraph structure (end)

\section{Background}
\label{sec:background}

On the way to exascale computing many new challenges are arising, and it is still unknown what problems we exactly have to expect. Consensus exists that the number of processes will increase by a factor of $10$ to $100$. At the same time it is anticipated that the Mean-Time-Between-Failure (MTBF) is decreasing. This is a direct effect of the increased number of processes, since studies predict that the MTBF is proportional to the number of processes~\cite{stearley:2010:redundant,schroeder:2007:failures}.
In addition systems reliability might become increasingly important because one approach to increase energy efficiency is to lower the core-voltage, which in turn leads to an increasing probability of soft faults (i.e. bit-flips) \cite{Lammers:2010:TEO}.

\subsection{Faults and Failures}
\label{sub:faults_and_failures}
Elliot et al.~\cite{Elliott:2014:ETI} introduce a widely used terminology and taxonomy to differentiate between fault types. These range from occasional or recurrent bit-flips which can have no effect or may lead to permanent faulty computation, up to losing complete nodes, and anything in between. For this work we categorise them roughly into two types: \emph{Hard failures} lead to the crash of a process, and \emph{soft failures} possibly lead to unexpected behaviour but without interrupting computation or communication. Nevertheless the obtained results after a soft failure may be faulty. In the following we shortly describe the characteristics of these failures. We do not differentiate between \emph{failure} and \emph{fault}.

\paragraph{Soft failure} % (fold)
\label{par:soft_failure}

We categorise a failure as a soft failure if afterwards the process is still capable of throwing an exception in some way. They can occur either directly due to a C++ runtime error, a numerical failure within an algorithm (like division by zero) or more generally a detected misbehaviour (e.g. solver divergence) and a related user-defined exception. In addition, the process must be able to communicate this exception afterwards. We will not further categorise or differentiate this type of failure. Neither we will talk about detection mechanism or possibilities to repair the effect of such failures. This is a user-level task and thus coherent with the idea behind the proposed ULFM extension.
We are interested in providing additional functionality for the user to handle such circumstances in a problem-specific and thus more efficient fashion, rather than a black-box solution like global checkpointing.
 % paragraph soft_failure (end)

\paragraph{Hard failure} % (fold)
\label{par:hard_failure}
A hard failure on the other hand leads to the loss of a part of a communicator, i.e. a process or a whole node. Within an MPI communication this can result in a deadlock due to open MPI requests. These failures are a main motivation behind the design of the ULFM extension~\cite{Bosilca:2016:FDP:3014904.3014941}. % which is essential to tolerate such faults in an application as described later in section \ref{sub:ulfm}.
If a hard failure occurs it is not straight forward to continue the computation. The default way to handle such faults is a rollback to a previous checkpoint, which will be more and more expensive with increasing parallelism not only because of recomputation but also because of communication~\cite{stearley:2010:redundant,Elliott:2012:CPR,Ferreira:2011:EVP,Fiala:2012:DAC}.  In addition the communicator has to be re-established with replacement processes, or  the application has to be repartitioned and/or load-balanced.
 % paragraph hard_failure (end)

\subsection{ULFM}
\label{sub:ulfm}
\emph{User Level Failure Mitigation} (ULFM) is proposed to be \lq a set of MPI interface extensions to enable MPI programs to restore MPI communication capabilities disabled by failures\rq\footnote{\url{http://fault-tolerance.org/downloads/20161115-tutorialSC16-handson.pdf}, slide 7}.
If a hard failure occurs in current versions of OpenMPI and MPICH, the runtime tries to terminate all processes and ends the computation.
The idea of the ULFM proposal is instead to return an error code to the user, which enables to define an approach to repair the computation, e.g. by freeing all other processes and the faulty communicator, followed by setting up a new communicator. Alternatively it is an option to shrink the communicator so that computation can be continued with less participants. We emphasise again that the actual reaction is problem-specific. To this end, ULFM proposes a set of new essential MPI functions, for instance:
\begin{itemize}
\item \lstinline!MPI_Comm_revoke!\\
  This function signals the revocation of the communicator to all ranks; further MPI calls (except of the two following) within the communicator will fail with an error code of class \lstinline!MPI_ERR_COMM_REVOKED!.
\item \lstinline!MPI_Comm_agree!\\
  This function provides functionality for agreeing on further proceeding after a failure between ranks within a communicator. A bit-wise \lstinline!AND! operation is performed over an integer.
\item \lstinline!MPI_Comm_shrink!\\
  A new communicator is created excluding all failed ranks.
\item Hard-failure detection\\
  Hard-failures are detected by ULFM. In that case all communication involving failed ranks is terminated with an error of class \lstinline!MPI_ERR_PROC_FAILURE! or \lstinline!MPI_ERR_PROC_FAILURE_PENDING!.
\end{itemize}
Several other features especially for file I/O and one-sided communication are provided by ULFM, and we refer to the current specification for details\footnote{\url{http://fault-tolerance.org/ulfm/ulfm-specification}}.

ULFM is not included in the MPI-3 standard but is proposed for MPI 4.0\footnote{\url{http://mpi-forum.org/mpi-40}}. A prototype implementation in OpenMPI exists, but is based on the outdated version 1.7.x of OpenMPI. This makes it hard to deploy it on current HPC systems and to make it available for a large user-base before it is finally integrated into the standard. A dedicated version of a ULFM-extended OpenMPI implementation is available on Edison, a large-scale Cray XC30 machine \footnote{\url{http://fault-tolerance.org/2017/03/26/running-on-edison}}. This indicates that it could become available in more production environments in the near future.

Some first implementations within MPICH~\cite{Bland:2015:ULMFMPICH} exist, but they are unoptimised yet and more or less only a proof of concept. In particular, it is not possible to shrink a revoked communicator\footnote{https://github.com/pmodels/mpich/issues/2198}, which is required in our ULFM-based implementation. Therefore we cannot present  results with MPICH version 3.3a2.

\section{Interface and technical details}
This section describes the user interface of our proposal. It implements the future
paradigm, which was introduced to C++ with C++11 to handle
asynchronous tasks. Furthermore it uses exceptions to indicate errors
like it is advocated by the C++ standard. Early experiments (not covered in this paper) indicate that our approach can already be used to define algorithm-specific LFLR techniques to increase the
fault-tolerance of algorithms.

For the implementation of the user interface we distinguish two cases:
with and without ULFM support of the underlying MPI implementation. We
refer the non-ULFM implementation as the \emph{Black-Channel}
approach since we create an additional communicator for error
communication, which is not used in the fault-free scenario. In this
case it is only possible to detect soft faults. If the ULFM extension is available, the interface will adapt and can
detect hard faults as well.

The code is available at \url{https://gitlab.dune-project.org/christi/test-mpi-exceptions} and contains also a range of small examples with exception propagation.

% In the ULFM case it is possible to detect hard failures
% and the error propagation strategy is implemented by the MPI
% realization, which may be customized on the hardware for performance
% reasons.
\subsection{User interface}\label{sub:userinterface}

\begin{figure*}
  \centering
  \includegraphics[width=0.8\textwidth]{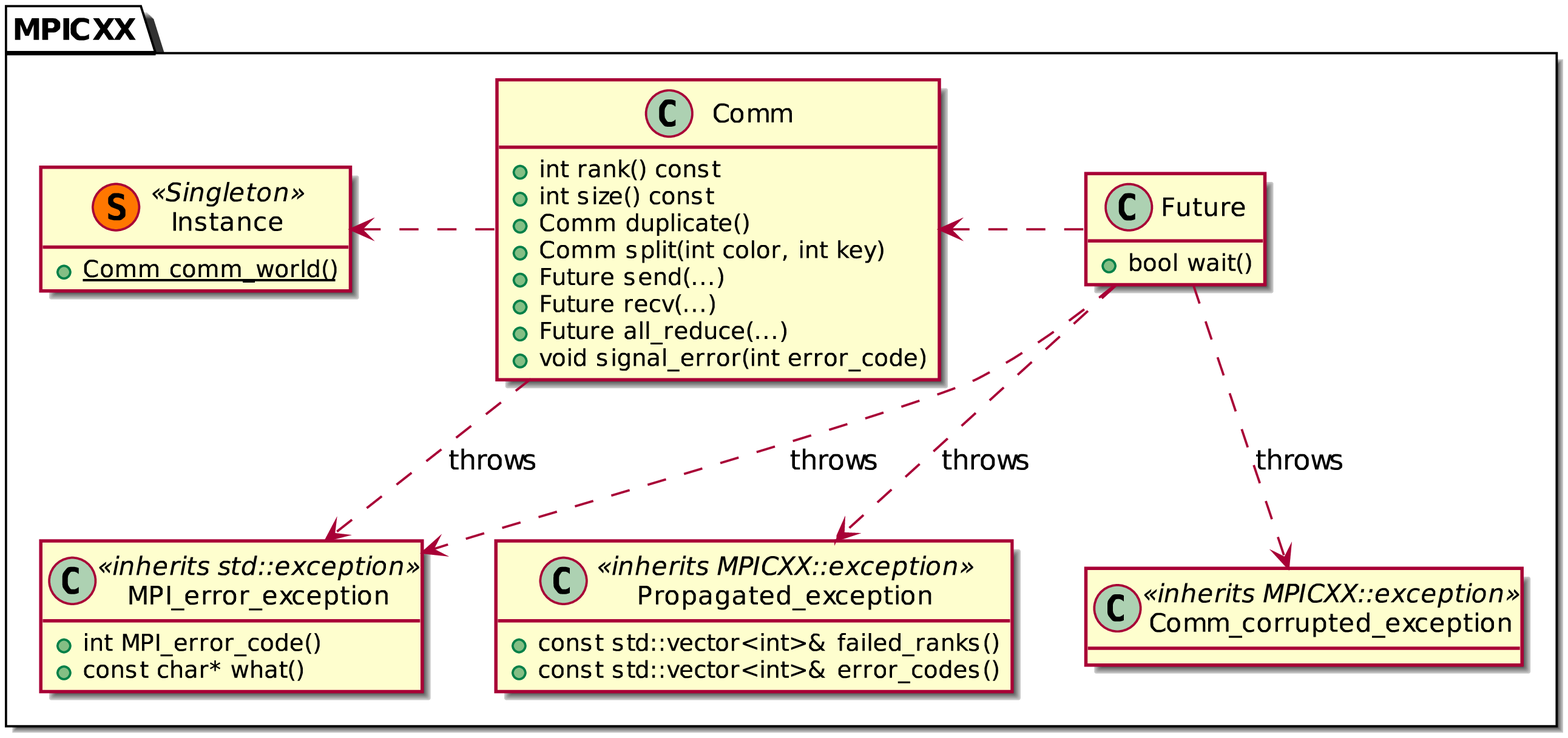}
  \caption{Class diagram of the user interface}
  \label{fig:user_interface}
\end{figure*}

Figure~\ref{fig:user_interface} shows a class diagram of the user
interface, and we describe the semantic of each class in the following paragraphs. Listing~\ref{lst:example} shows an example of using the user interface.

\paragraph{Instance}
Every MPI program needs to call \lstinline!MPI_Init!  at the beginning
of the program and
\lstinline!MPI_Finalize! at the end. We implement this into a singleton class
\lstinline!Instance! to ensure the proper structure of the program. The constructor checks if MPI is already initialised, if it is not the case \lstinline!MPI_Init! is called. \lstinline!MPI_Finalize! is only called in the destructor if \lstinline!MPI_Init! was called in the constructor of the respective object. The \lstinline!Instance! class also provides access to the \lstinline!comm_world! communicator.

\paragraph{Comm}
The class \lstinline!Comm! manages a communicator and provides
functions to duplicate the communicator, to generate new
sub-communicators and to issue communication calls
like \lstinline!send! and \lstinline!recv!. Collective communication is also feasible but might lead to a memory leak in the failure case (cf. section \ref{subsec:issues}). We exemplarily implemented the \lstinline!all_reduce! functionality but the class is easily extendable to every non-blocking communication method. Furthermore
\lstinline!rank!  number and \lstinline!size! of the communicator can
be determined by calling the respective methods. Instances of this
class cannot be copied, since this class represents a one-to-one
relation to MPI communicators. For duplication the interface provides
a dedicated \lstinline!duplicate! method. Intracommunicators are not
supported yet.

\paragraph{Future}
Communication requests are wrapped in the class
\lstinline!Future! which implements the asynchronous
programming concept of using \textit{future-type objects}. An non-blocking communication can
be initiated by calling the respective methods of the \lstinline!Comm!
object, which returns a \lstinline!Future!  object. The user
calls the method \lstinline!wait! to ensure that the communication is completed (i.e. the buffer of the communication can be reused). The \lstinline!wait! method may
throw one of the following exceptions:

\paragraph{Propagated}
One rank can signal an error to all remote ranks by calling the method
\lstinline!signal_error!, which takes an error code as an argument. In
this case all remote ranks throw an exception of type
\lstinline!Propagated_exception!, when they call \lstinline!wait!  of
a \lstinline!Future! object or if they are already waiting. The rank itself throws a \lstinline!Propagated_exception!
within the method \lstinline!signal_error!. The \lstinline!Propagated_exception! objects contains
information about which ranks (possibly several) have signaled an error
and with which error code. Reacting to these exceptions does not require to revoke and set up a new communicator.

\paragraph{Corrupted communicator}
The \lstinline!Comm! object detects in the destructor whether it gets
destructed during stack unwinding due to a thrown exception by
using \lstinline!std::uncaught_exception!. This incident is
interpreted as an unrecoverable error within the communicator. It is
propagated to all remote ranks, which will throw a
\lstinline!Corrupted_comm_exception! when they call \lstinline!wait!
or are already waiting on a \lstinline!Future! object. This exceptions
should not be caught within the scope of the \lstinline!Comm! object
to ensure a consistent state on all ranks.

\paragraph{MPI errors}
In the case of any MPI error that cannot be assigned to one of the
previous exceptions we throw an \lstinline!MPI_error_exception!. It
inherits from \lstinline!std::exception! and contains the respective
error code.

{
\begin{lstlisting}[float=*, caption=Minimal user interface (2 processors), label=lst:example,basicstyle=\small]
// initializing mpi via a singleton class
const MPICXX::Instance& mpi = MPICXX::initialize(argc, argv);
try { // try-catch: corrupted communicator exceptions
    MPICXX::Comm& comm = mpi.comm_world(); // get a reference to the mpi communicator
    try { // try-catch: remote/propagated exceptions
        try { // try-catch: local exceptions
            int answer = 0;
            MPICXX::Future f; // creating a future object to store the communication
            if (comm.rank() == 0) { // rank 0
                answer = 42;
                // set up a send from rank 0 to rank 1
                f = comm.send(&answer, 1, MPI_INT, 1);
            }
            if (comm.rank() == 1) { // rank 1
                // set up a receive on rank 1 from rank 0
                f = comm.recv(&answer, 1, MPI_INT, 0);
            }
            f.wait();  // wait for communication to be finished
        } catch(std::exception& e) {  // catch local exception
            comm.signal_error(666);  // propagate an exception with code 666 to all ranks
        }
    } catch(MPICXX::Propagated_exception &e) { // catch a remote/propagated exception
        // initiate a recovery process: e.g. if a skewed Krylov basis is detected locally
        // a global restart with the current approximation can be initiated
        // further use cases are mentioned in the introduction
    }
} catch(MPICXX::Comm_corrupted_exception &e) { // catch a corrupted communicator exception
    // repair communicator, initiate rollback, ...
    // see Section II.B ULFM for further information
}
\end{lstlisting}}

\subsection{Black channel implementation}
We now present the implementation based on the MPI-3.0 standard
(without ULFM). The constructor of the \lstinline!Comm!
object duplicates the MPI communicator by calling
\lstinline!MPI_Comm_dup!. The new communicator is called
\lstinline!comm_err! and is stored in the \lstinline!Comm! object. It
is used for failure related communication. In \lstinline!comm_err! we
create a non-blocking receive operation via \lstinline!MPI_Irecv! and
store the pending request in \lstinline!err_req!. The duplication of
the communicator is made to not block a communication tag.

The function \lstinline!signal_error! issues a matching
\lstinline!MPI_Issend! for \lstinline!err_req! to all other ranks and
cancels its own \lstinline!err_req!. It uses the non-blocking
operation since it is possible that two ranks simultaneously
propagate errors: In that case a blocking operation may deadlock since
it is not ensured that there is a matching \lstinline!recv!. Once all
error messages have been send or a rank receives an error message, it
calls \lstinline!MPI_Barrier! to wait for all ranks being in the
error state. When all ranks reach the barrier, the propagating ranks
cancel the pending send requests, which are the send requests to the
ranks that got signaled by another rank. Then all ranks perform an
\lstinline!MPI_Allreduce! operation with an \lstinline!MPI_BAND!
operator to determine if the communicator is corrupted,
i.e. \lstinline!signal_error! was called by the destructor of
\lstinline!Comm! during stack unwinding. If the call results positive,
all ranks throw a \lstinline!Comm_corrupted_exception!, otherwise the
following algorithm determines the failed ranks and
respective error codes:

\paragraph{Determine failed ranks and codes}
Once one or more errors have been signaled, all ranks throw a
\lstinline!Propagated_exception! and all information is propagated to all ranks. For that, we do an \lstinline!MPI_Scan! with the operation
\lstinline!MPI_SUM!, where failed ranks participate with a $1$ and
non-failed ranks with a $0$. This assigns every failed node an
\lstinline!index!. The number of failed nodes is then propagated by an
\lstinline!MPI_Bcast! of the last rank (i.e. rank
\lstinline!size!$-1$). Now all ranks allocate memory for the rank
numbers and error codes of the failed ranks and initialise it with
zeros. The failed ranks write their rank number and error code to this array with respect to their \lstinline!index!. Finally an
\lstinline!MPI_Allreduce!  with \lstinline!MPI_MAX! is performed to
propagate all the information.

\paragraph{Future}
During computations the user initiates non-blocking requests by
calling the \lstinline!send! or \lstinline!recv! method of the
\lstinline!Comm!  object. The respective requests are stored in the
\lstinline!request!  field of a \lstinline!Future! object. Instead of
just waiting for the \lstinline!send!/\lstinline!recv! request to
finish in the \lstinline!wait! method, \lstinline!MPI_Waitany! is used
that waits for either the \lstinline!request! or the error requests
\lstinline!err_req! to complete. It is possible that
\lstinline!MPI_Waitany! completes \lstinline!request! while an error
was signaled as well. Therefore, if \lstinline!MPI_Waitany! completes
\lstinline!request!, the method uses \lstinline!MPI_Test! to check whether
an error was signaled. If no error code is received, the program
continues its computations. Otherwise an error code is
received, the above described algorithm is executed to handle the error
and throw the appropriate exception.

\paragraph{Corrupted communicator}
The destructor of the \lstinline!Comm! class checks whether the
object is deconstructed due to a thrown exception. If this is the
case, \lstinline!signal_error! is called. At the following
\lstinline!MPI_Allreduce! this rank participates with a $0$, indicating that the
communicator is corrupted. All other ranks will throw a
\lstinline!Comm_corrupted! exception.

\paragraph{Preclusion of deadlocks}
This approach precludes deadlocks that are caused by thrown
exceptions, since either the program executes successfully, i.e. all
\lstinline!request!s are completed by the \lstinline!MPI_Waitany!
method or, if an exception is thrown, an error is signaled and the
\lstinline!MPI_Waitany! will return with an error and throw a
\lstinline!Comm_corrupted_exception!. Furthermore, in erroneous cases,
the execution path of the ranks can be synchronised with the
\lstinline!signal_error! method.

\subsection{ULFM adoption}\label{sub:ulfm_adoption}

Until ULFM is available on HPC clusters, the presented Black-Channel
approach can be used to develop fault-tolerant programs using the interface described in Section~\ref{sub:userinterface}. However, as soon as
ULFM is available, it constitutes the proper tool to handle failures. As mentioned earlier, ULFM enables the detection of
hard failures and even communicates this to other ranks, which our proof-of-concept Black-Channel cannot. Therefore
we adapt our implementation to use ULFM features if available. This makes it possible to increase the functionality of user-level code written against our interface in the future, without changing the general strategy to react to erroneous behaviour.

If ULFM is available, the \lstinline!wait! method of the
\lstinline!Future! invokes an \lstinline!MPI_Wait!, instead of the
\lstinline!MPI_Waitany!, and checks the return code. This
\lstinline!MPI_Wait! call returns with the error code
\lstinline!MPI_ERR_REVOKED!, if any rank has called
\lstinline!MPI_Comm_revoke!. Also the additional communicator \lstinline!err_comm! and the
pending \lstinline!MPI_Irecv! requests are not necessary any more.

After the communicator is revoked the function
\lstinline!MPI_Comm_agree! is used to determine whether the
communicator is corrupted or an error code is signaled. If the
communicator is corrupted a \lstinline!Comm_corrupted_exception! is
thrown, otherwise \lstinline!MPI_Comm_shrink! is called to obtain a valid
communicator. Then we proceed with the same algorithm like in the
Black-Channel case to propagate the rank numbers and error codes of
the failed ranks.

There are three cases in which the communicator is revoked. The first
case is the call of the method \lstinline!signal_error!. The following
\lstinline!MPI_Comm_agree! proceeds with $1$ on all ranks, indicating
that the communicator is not corrupted.  The other cases are when the
communicator object is deconstructed during
stack unwinding caused by a thrown exception, or
that an MPI call returns \lstinline!MPI_ERR_PROC_FAILED! or
\lstinline!MPI_ERR_PROC_FAILED_PENDING!. The latter implies a hard
failure of a node or rank. In these cases the respective ranks participate
with $0$ at the following \lstinline!MPI_Comm_agree!, indicating that
the communicator is corrupted and a
\lstinline!Comm_corrupted_exception! is thrown on all ranks.

\section{Validation and Discussion}

\subsection{Validation}
We tested both implementations on PALMA, the HPC cluster of the University of Münster. We use 12 nodes and 48 nodes, with 12 processes each, i.e. 144 and 576 ranks, respectively. Every node consists of two hexacore Intel Westmere processors, and the nodes are connected by QDR InfiniBand. IntelMPI (version 5.1.3) and OpenMPI (version 1.8.4) use the RDMA protocol on the interconnect. The ULFM variant of OpenMPI (based on OpenMPI 1.7.1) does not support this, thus TCP/IP over InfiniBand is used. This drawback of the OpenMPI-ULFM implementation unfortunately affects the latency of the system. For reference, we also show timings for the newer OpenMPI version and TCP/IP over InfiniBand. In the OSU Benchmark (osu\_barrier) \cite{panda2013osu}, the ULFM instance is 35 times slower than IntelMPI, and 6 times slower than the standard OpenMPI installation, see Table~\ref{tab:benchmark} for details.

\begin{table}[htb]
  \centering
  \begin{tabular}{c|c|c|c}
    IntelMPI & OpenMPI & OpenMPI (tcp) & OpenMPI-ULFM\\
    \hline
    $16.7\mu s$ & $97.3\mu s$ & $502.6\mu s$ & $585.5\mu s$
  \end{tabular}
  \caption{Average latency of the OSU Benchmark (osu\_barrier, 1000 iterations) on PALMA.}
  \label{tab:benchmark}
\end{table}

\begin{figure}[htb]
  \centering
  \begin{subfigure}{0.45\textwidth}
    \includegraphics[width=\textwidth]{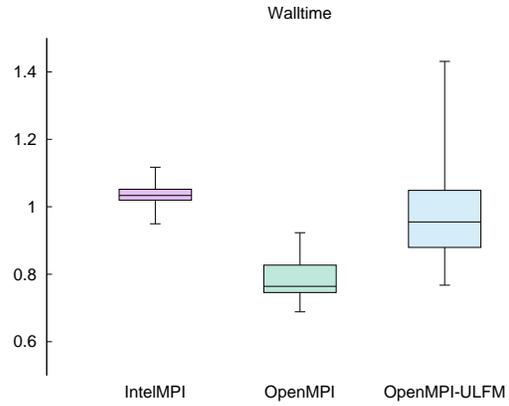}
    \caption{Propagating an error on 12 nodes with 12 processes each.}
  \end{subfigure}
  \qquad
  \begin{subfigure}{0.45\textwidth}
    \includegraphics[width=\textwidth]{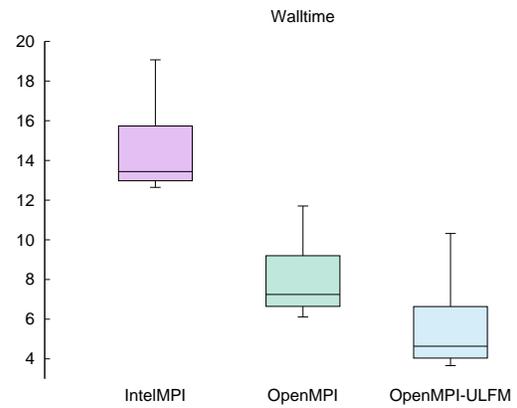}
    \caption{Propagating an error on 48 nodes with 12 processes each.}
  \end{subfigure}
  \caption{Duration of the propagation}
  \label{fig:time}
\end{figure}

In addition to testing the functionality, we measure the time that is needed to propagate an exception. We measure the time for duplicating \lstinline!comm_world!, propagating an exception from rank 0 and cleaning up the duplicated communicator, i.e., we simultaneously measure the overhead and propagation time. For the Black-Channel approach this results in two calls of \lstinline!MPI_Comm_dup! while with ULFM only one call is necessary. We repeat this test 1000 times for the MPI implementations available to us. Figure~\ref{fig:time} shows boxplots over the duration measured on the root rank in
milliseconds. While the Black-Channel approach is competitive at 12 nodes, we observe that on 48 nodes the Black-Channel approach (for both libraries)
is slower than the ULFM implementation, as it is not as optimised as the algorithms used in ULFM. However, the OpenMPI implementation of ULFM is based on an older version of OpenMPI and is not optimised for performance yet, thus further speed-up of our propagation strategy can be expected, once ULFM is integrated into the standard.

\subsection{Issues}
\label{subsec:issues}
The main issue with the Black-Channel approach is that it only works
properly for point-to-point communication. If we invoke a non-blocking
collective communication, we cannot call \lstinline!MPI_Cancel! for these
requests. The standard states explicitly:
\begin{quote}{MPI Standard 3.1 \cite{mpi:2015}, page 197}
  It is erroneous to call \lstinline!MPI_REQUEST_FREE! or
  \lstinline!MPI_CANCEL! for a request associated with a nonblocking
  collective operation. \texttt{[...]}

  \emph{Rationale.}  Freeing an active nonblocking collective request
  could cause similar problems as discussed for point-to-point
  requests (see Section 3.7.3). Cancelling a request is not supported
  because the semantics of this operation are not well-defined.
  \emph{(End of rationale.)}
\end{quote}

This implies that all buffers involved in
the non-blocking collective communication should be valid until the request
finishes (which will, in many cases, never happen). This extra memory
and the state information of the actual request in some internal MPI
data structures might be negligible in many cases,
e.g. \lstinline!MPI_Ireduce!, but is becoming a problem when gather/scatter operations on a large communicator are called. On the other hand
for really large scale computations, collective
communication should be avoided anyway and one should work on small subcommunicators, as the cost
rises with the number of ranks. Then also the (unavoidable) memory
leak is small.  The real issue arises when long time computations are
considered, as the communicator (according to the current MPI
standard) cannot be freed (internally) as long as there are open
requests on this communicator.

The problem is solved once the proposed ULFM extension is
included in the MPI standard: This deprecates our Black-Channel workaround, and supports to revoke a
communicator and cancel all open requests.
% MPI-ULFM introduces
% \lstinline!MPI_Comm_revoke! to invalidate an existing communicator and
% cancel all associated requests. The requests will return with an
% appropriate failure.

An issue with the Black-Channel approach is that the propagation of the error needs at least
$n-1$ point-to-point communications that are initiated by a single
rank. This might imply problems if many ranks are used and errors
occur often. However, optimising the performance of the error
propagation strategy needs knowledge of the network topology and is
therefore not discussed further in this paper. ULFM might also be the solution, since the
propagation strategy in \lstinline!MPI_Comm_revoke! is implementation
dependent, and MPI implementors are known for providing extremely efficient realisations of the MPI standard.

\bibliographystyle{IEEEtran}
\bibliography{IEEEabrv,references}

\end{document}